\documentclass[10pt, twocolumn, notitlepage, superscriptaddress, prl, longbibliography]{revtex4-2}

\usepackage{here}
\usepackage{amsmath}         
\usepackage{graphicx}
\usepackage{braket}         
\usepackage{lettrine}
\usepackage{xcolor}
\usepackage{times}
\usepackage{hhline}
\usepackage{tabularx}
\usepackage{calc}
\usepackage{soul}

\usepackage[colorlinks=true,linkcolor=blue, citecolor=blue]{hyperref}%

\begin{document}

\title{Long-Range Orbital Magnetoelectric Torque in Ferromagnets}    

\author{Dongwook Go}
\email{d.go@fz-juelich.de}
\affiliation{Peter Gr\"unberg Institut and Institute for Advanced Simulation, Forschungszentrum J\"ulich and JARA, 52425 J\"ulich, Germany \looseness=-1}
\affiliation{Institute of Physics, Johannes Gutenberg University Mainz, 55099 Mainz, Germany}

\author{Daegeun Jo}
\affiliation{Department of Physics, Pohang University of Science and Technology, Pohang 37673, Korea \looseness=-1}

\author{Kyoung-Whan Kim}
\affiliation{Center for Spintronics, Korea Institute of Science and Technology, Seoul 02792, Korea}

\author{Soogil Lee}
\affiliation{Department of Materials Science and Engineering and KI for Nanocentury, KAIST, Daejeon 34141, Korea}

\author{Min-Gu Kang}
\affiliation{Department of Materials Science and Engineering and KI for Nanocentury, KAIST, Daejeon 34141, Korea}

\author{Byong-Guk Park}
\affiliation{Department of Materials Science and Engineering and KI for Nanocentury, KAIST, Daejeon 34141, Korea}

\author{Stefan Bl\"ugel}
\affiliation{Peter Gr\"unberg Institut and Institute for Advanced Simulation, Forschungszentrum J\"ulich and JARA, 52425 J\"ulich, Germany \looseness=-1}

\author{Hyun-Woo Lee}
\affiliation{Department of Physics, Pohang University of Science and Technology, Pohang 37673, Korea \looseness=-1}

\author{Yuriy Mokrousov}
\affiliation{Peter Gr\"unberg Institut and Institute for Advanced Simulation, Forschungszentrum J\"ulich and JARA, 52425 J\"ulich, Germany \looseness=-1}
\affiliation{Institute of Physics, Johannes Gutenberg University Mainz, 55099 Mainz, Germany}

\begin{abstract}
While it is often assumed that the orbital response is suppressed and short-ranged due to strong crystal field potential and orbital quenching, we show that the orbital magnetoelectric response can be remarkably long-ranged in ferromagnets. In a bilayer consisting of a nonmagnet and a ferromagnet, spin injection from the interface results in spin accumulation and torque in the ferromagnet, which rapidly oscillate and decay by spin dephasing. In contrast, we find that even when an external electric field is applied only on the nonmagnet, we find substantially long-ranged orbital magnetoelectric response in the FM, which can go far beyond the spin dephasing length. This unusual feature is attributed to nearly degenerate orbital characters imposed by the crystal symmetry, which form hotspots for the intrinsic orbital response. Because only the states near the hotspots contribute dominantly, the induced OAM does not exhibit destructive interference among states with different momentum as in the case of the spin dephasing. This gives rise to a distinct type of orbital torque on the magnetization, increasing with the thickness of the ferromagnet. Such behavior may serve as critical long-sought evidence of orbital transport to be directly tested in experiments. Our findings open the possibility of using long-range orbital magnetoelectric effect in orbitronic device applications. 
\end{abstract}

\date{\today}                 
\maketitle	      


Over the past decades, it has been realized that electronic current can carry information not only about charge, but also spin, orbital, and valley degrees of freedom, which generated separate fields of spintronics \cite{Wolf2001, Zutic2004, Fert2008, Bader2010, Hirohata2020}, orbitronics \cite{Bernevig2005, Go2017, Phong2019}, and valleytronics \cite{Schaibley2016, Vitale2018}. Fundamental studies of spin currents \cite{Sinova2015, Manchon2019} have led to practical applications of spintronic devices: e.g. spin torque switching for  magnetic memory devices \cite{Bhatti2017} and spin torque oscillator for high frequency generators and neuromorphic computing \cite{Chen2016, Torrejon2017, Grollier2020}.
Analogous to the spin current, recent studies have shown that the orbital current can be electrically generated via the orbital Hall effect (OHE) in various systems \cite{Bernevig2005, Tanaka2008, Kontani2009, Go2018, Jo2018, Salemi2020, Canonico2020a, Canonico2020b, Bhowal20020b, Bhowal20020b, Cysne2021, Sahu2021, Bhowal2021}. The OHE naturally explains the variation in the magnitude and sign of the spin Hall effect in terms of the correlation between the spin and orbital variables \cite{Tanaka2008, Kontani2009, Go2018}. Moreover, the orbital current can interact with the magnetic moment through spin-orbit coupling (SOC) and induce magnetization dynamics \cite{Go2020}. This finding has not only triggered ideas in orbitronics but has attracted interest from the spintronics side since the efficiency of the OHE is generally much higher than that of the spin Hall effect \cite{Tanaka2008, Kontani2009, Go2018, Jo2018} and the SOC is not required for the generation of  orbital currents \cite{Jo2018}. This may help to overcome restrictions on material choices for spintronic devices beyond heavy elements which are employed to achieve large spin Hall effect.

One of the biggest challenges in detecting the orbital current is its resemblance to the spin current, i.e. both orbital angular momentum (OAM) and spin transform in the same way under symmetry operations, which makes it hard to distinguish them. Thus, most experiments performed so far relied on quantitative theoretical predictions that the spin contribution is significantly smaller than the orbital contribution in light metals \cite{Zheng2020, Kim2021, Choi2021, Ding2022, Liao2022}, their signs are the opposite \cite{Tazaki2020, Lee2021b, Hayashi2022}, and the orbital is efficiently converted into the spin by strong SOC of heavy metals \cite{Ding2020, Lee2021a, Hu2022}. This motivates us to search for a unique fingerprint of the orbital excitation, which is derived from the microscopic nature and is qualitatively different from the spin excitation. 

We emphasize that the OAM and spin are fundamentally different objects. While spin, to a large degree, is a good quantum number unless the SOC is strong, the OAM is generally not conserved even when the SOC is weak as it strongly interacts with the lattice~\cite{Haney2010, Go2020b, Han2022}. Thus, the orbital excitation has long been regarded fragile and short-ranged. However, recent experiments found evidence of strong \emph{orbital MEC} $-$ the OAM induced by an external electric field \cite{Choi2021, Tazaki2020, Kim2021, Hu2022, Ding2022, Hayashi2022, Liao2022}. In particular, Refs. \cite{Ding2022, Hayashi2022, Liao2022} suggest a highly nonlocal nature of the induced OAM, which affects magnetic moments even $\gtrsim 20\ \mathrm{nm}$ away from the interface. This implies the presence of a robust mechanism for the orbital MEC over very long distance, which is not only exotic, but may also serve as a qualitatively distinct feature when compared to spin response.

\begin{figure}[t!]
\includegraphics[angle=0, width=0.45\textwidth]{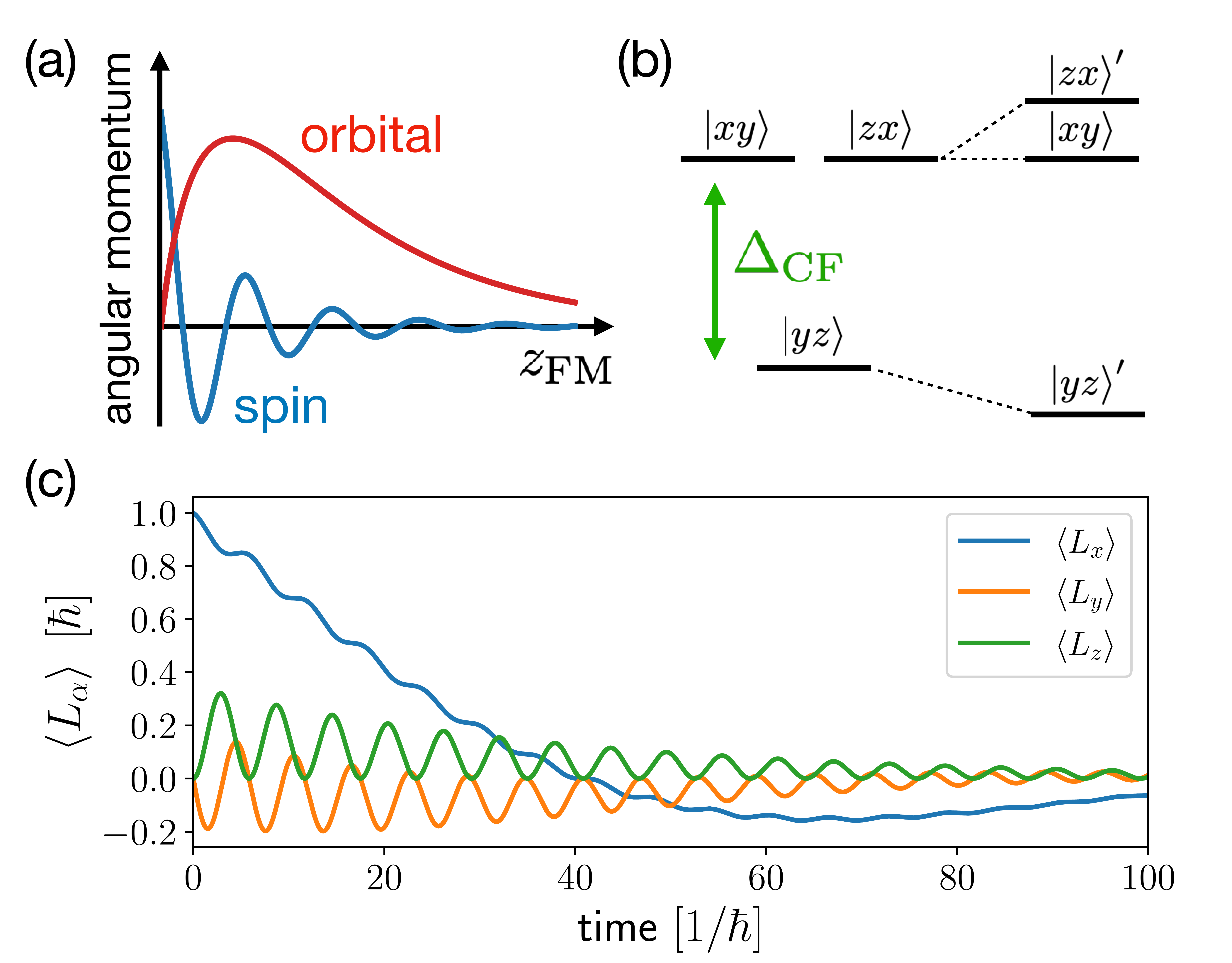}
\caption{
\label{fig:toy_model}
(a) Schematic illustration of the orbital (red) and spin (blue) response in the FM. While the spin oscillates and decays quickly, the OAM does not oscillate and decays monotonically. (b) Energy levels of $t_{2g}$ orbitals at $\mathbf{k}=(k_x,0,0)$. Note that $\ket{xy}$ and $\ket{zx}$ states are nearly degenerate and are separated from the $\ket{yz}$ state by the energy gap $\Delta_\mathrm{CF}$. (c) Such an energy level structure leads to slow evolution of $\langle L_x \rangle$ compared to those of $\langle L_y \rangle$ and $\langle L_z \rangle$.
}
\end{figure}

In this Letter, we unveil long-range nature of orbital magnetoelectric coupling (MEC) in ferromagnets (FMs) and the resulting torque on the magnetization. In a bilayer consisting of a nonmagnet and a FM, we find an external electric field applied on the nonmagnet can result in nonlocal orbital MEC and torque in the FM even far away from the interface. In contrast to spin response, which rapidly oscillates and decays, the OAM does not oscillate and propagate even longer distance as schematically shown in Fig.~\ref{fig:toy_model}(a). This is in stark contrast to a common expectation that the orbital transport is short-ranged because of the orbital quenching. We show that this originates in the hotspots where particular orbital characters are almost degenerate due to a crystal symmetry. At such hotspots, not only the OAM response is pronounced but also the dynamics is strongly suppressed, which works as an \emph{orbital trap}. Since the induced OAM interacts with the magnetization and exerts a torque, the long-range orbital MEC can be readily tested in experiments by measuring the dependence of the magnetic torque as a function of the FM thickness.


To provide an intuitive physical argument, we consider a toy model with $t_{2g}$ $d$-orbitals in cubic crystals. The energy level structure is schematically shown in Fig. \ref{fig:toy_model}(b) for states along $k_x$ axis. By a discrete rotation symmetry with respect to $x$-axis, $\ket{xy}$ and $\ket{zx}$ are degenerate, but $\ket{yz}$ is split by the crystal field of magnitude $\Delta_\mathrm{CF}$. In FMs, $\ket{xy}$ and $\ket{zx}$ are slightly split by a combined action of the spin exchange interaction and the SOC, which is much smaller than $\Delta_\mathrm{CF}$. Energy levels at arbitrary $\mathbf{k}$ points can be analogously understood, e.g. in terms of \emph{radial} and \emph{tangential} orbital characters with respect to the direction of $\mathbf{k}$ \cite{Go2018, Jo2018, Han2022}.

The orbital level structures imposed by the crystal potential has significant impact on the \emph{dynamics} of the OAM. We demonstrate this by explicitly solving the time-dependent Schr\"{o}dinger equation with the boundary condition $\ket{\psi (t=0)}=(-i\ket{xy}+\ket{zx})/\sqrt{2}$ for which $\langle L_x \rangle=\hbar$. The Hamiltonian is given by $\mathcal{H}_\mathrm{tot} = \mathcal{H}_\mathrm{CF} + J_\mathrm{orb} L_z$, which effectively describes the orbital degree of freedom in the FM with weak SOC when the magnetization is pointing along $z$. Here, $\mathcal{H}_\mathrm{CF}$ describes the crystal field splitting $\Delta_\mathrm{CF}$ between $\ket{xy}, \ket{zx}$ and $\ket{yz}$, and $J_\mathrm{orb}$ is an effective orbital Zeeman field that incorporates a combined action of the spin exchange interaction and the SOC. We also introduce a quasi-particle lifetime $\tau=2\hbar/\eta$, where $\hbar$ is the reduced Planck constant. For $\Delta_\mathrm{CF}=1.0$, $J_\mathrm{orb}=0.2$, and $\eta=0.05$ (dimensionless energy unit), the time-evolution of the OAM expectation value is shown in Fig.~\ref{fig:toy_model}(c). Interestingly, the dynamics of $\langle L_x \rangle$ is much slower than the dynamics of $\langle L_y \rangle$ and $\langle L_z\rangle$. The fast oscillation of $\langle L_y \rangle$ and $\langle L_z\rangle$ is due to large $\Delta_\mathrm{CF}$, leading to frequency $w \sim \Delta_\mathrm{CF}/\hbar$.

The above analysis implies that once $L_x$ is induced at $\mathbf{k}$-points near $k_x$ axis, its time-evolution is suppressed. The Kubo formula implies that the hotspot for a response of $L_x$ is also around $k_x$ axis where $\ket{zx}$ and $\ket{yz}$ are nearly degenerate. Therefore, these $\mathbf{k}$-points serve not only as \emph{hostpots} for the OAM response under an external perturbation but also as \emph{orbital traps} in terms of dynamics. Correspondingly, these features result in long-range response of the OAM. This is different from spin dephasing in FMs, which originate in fast precession of spins by strong exchange interaction and a destructive interference among states at different $\mathbf{k}$  \cite{Stiles2002, Slonczewski2002, Zhang2004, Zwierzycki2005}. 

\begin{figure}[t!]
\includegraphics[angle=0, width=0.5\textwidth]{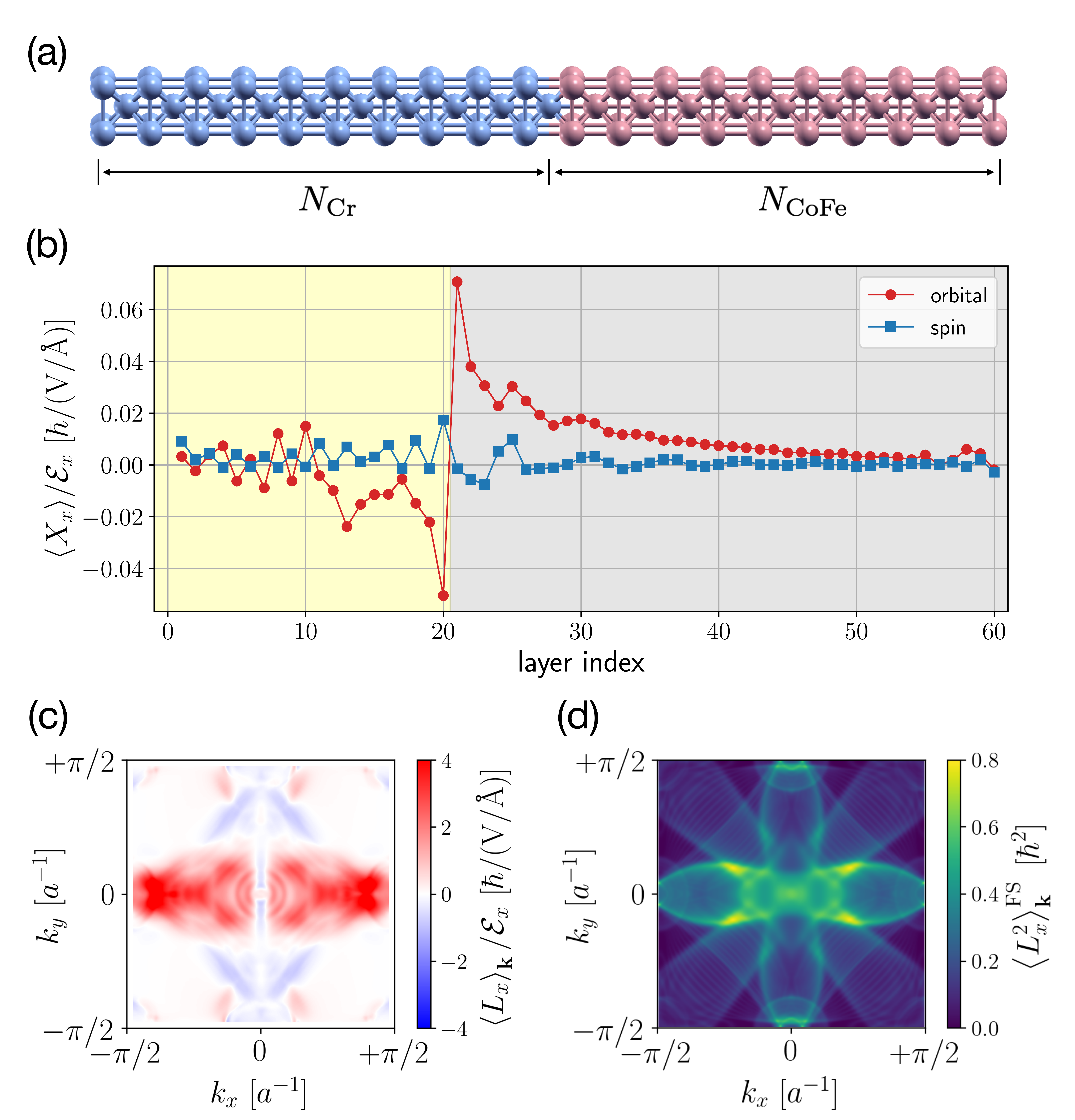}
\caption{
\label{fig:induced_OM}
(a) Structure of the Cr/CoFe bilayer, where Cr (light blue spheres) and CoFe (pink spheres) layers consist of $N_\mathrm{Cr}$ and $N_\mathrm{CoFe}$ atomic layers, respectively. We set $N_\mathrm{Cr}=20$ and $N_\mathrm{CoFe}=40$ in the calculation.
(b) Intrinsic responses of the $x$ component of OAM (red circle symbols) and spin (blue square symbols) by an external electric field applied on the Cr layers.
(c) $\mathbf{k}$-space distribution of the induced OAM in the CoFe layers. (d) The Fermi surface orbital texture $\langle L_x^2 \rangle_\mathbf{k}^\mathrm{FS}$ in equilibrium for the same CoFe layers.
}
\end{figure}

To demonstrate this in a real material, we perform numerical calculations based on the realistic tight-binding model constructed from first-principles, whose details can be found in~\cite{Supplementary}. We consider a Cr($N_\mathrm{Cr}$)/CoFe($N_\mathrm{CoFe}$) in body-centered cubic (001) stacking along $z$, where the numbers in the parenthesis indicate the number of atomic layers [Fig.~\ref{fig:induced_OM}(a)]. We assume that CoFe is perpendicularly magnetized, and Cr is nonmagnetic. We calculate \emph{intrinsic} responses of the OAM and spin under an external electric field  along $x$ by the Kubo formula. We assume that the electric field is applied only to Cr layers to investigate the consequences of the orbital injection by the OHE in Cr. We emphasize that Cr exhibits gigantic OHE \cite{Jo2018}, whose evidences have been found in experiments~\cite{Lee2021a}.



As the $y$-component of the OAM is injected by the OHE into CoFe, its precession with respect to the magnetic moment generates $\langle L_x \rangle$. Figure~\ref{fig:induced_OM}(b) shows responses of $\langle L_x \rangle$ and $\langle S_x \rangle$ in each layer of Cr(20)/CoFe(40). Although the electric field is applied only on the Cr layers, we find gigantic orbital response in CoFe layers, revealing nonlocal nature of the orbital MEC. The induced OAM does not oscillate but decays monotonically, propagating for up to $\sim 30$ atomic layers. In contrast, the induced spin shows an oscillatory decay behavior, and the magnitude is much smaller than that of the OAM response. The numerical result agrees with the qualitative picture of Fig.~\ref{fig:toy_model}(a). 



In order to confirm that the long-range orbital MEC is due to the hotspot, we calculate induced OAM $\langle L_x \rangle_\mathbf{k}$ for each $\mathbf{k}$-point. Figure~\ref{fig:induced_OM}(c) shows the distribution of $\langle L_x \rangle_\mathbf{k}/\mathcal{E}_x$. It clearly shows that the hotspot is located along $k_x$ axis, where a nearly degenerate orbital character of the states facilitates a large response of $L_x$. This can be characterized by the expectation value of $L_x^2$ in equilibrium, which is shown in Fig.~\ref{fig:induced_OM}(d) for the states at the Fermi surface. In accord with the hotspot picture discussed above, Figs.~\ref{fig:induced_OM}(c) and \ref{fig:induced_OM}(d) confirm that the microscopic origin of the orbital magnetoelectric response is due to the degenerate structure of the orbital character.
We remark that the hotspot-origin is well-known for many intrinsic response phenomena such as anomalous Hall effect \cite{Yao2004, Wang2006}. However, in most cases, the hotspots are  accidental degeneracies rather than degeneracies imposed by the symmetry. 

\begin{figure}[t!]
\includegraphics[angle=0, width=0.46\textwidth]{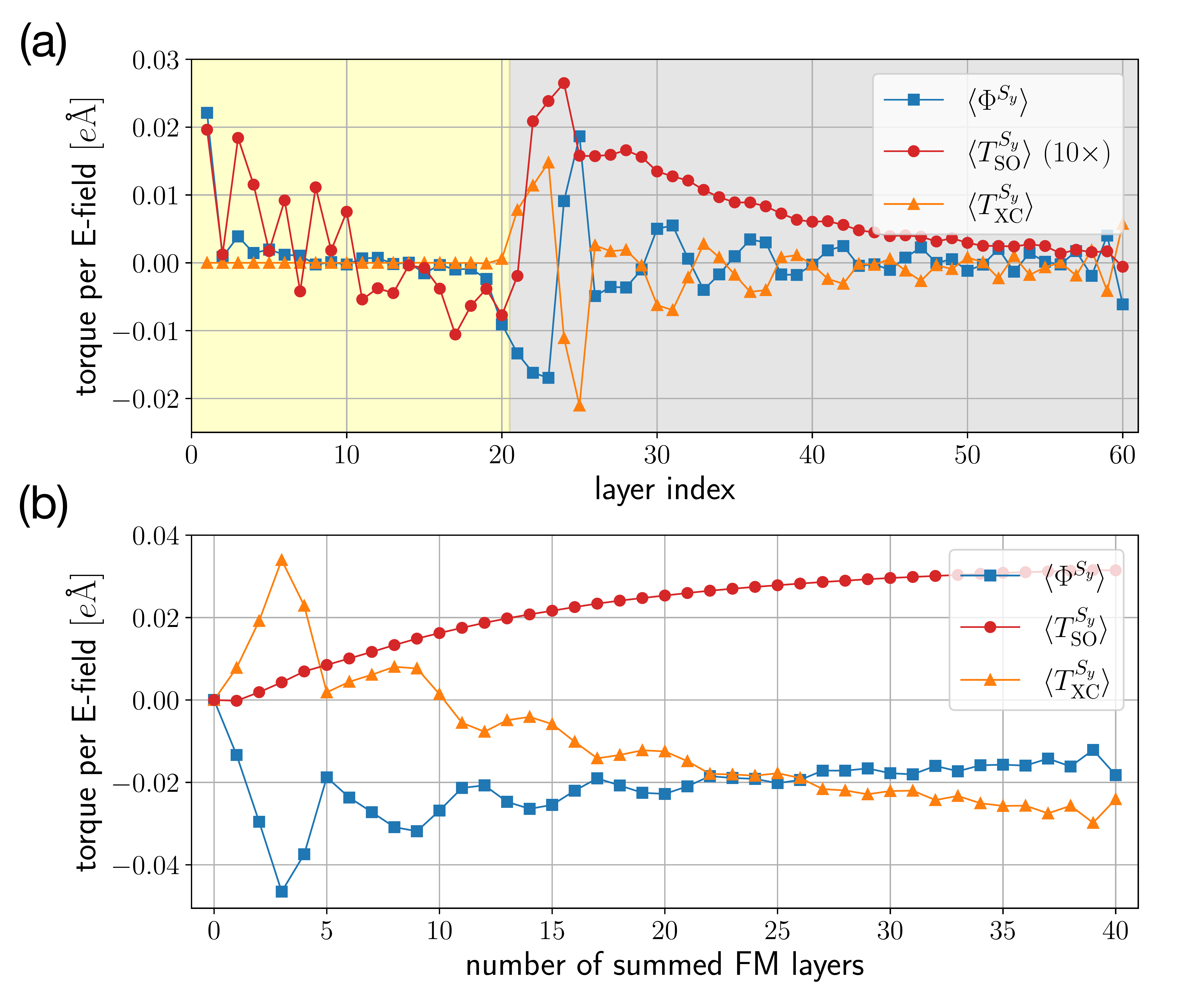}
\caption{
\label{fig:torque_analysis}
(a) The intrinsic responses for the spin flux $\langle \Phi^{S_y} \rangle$ (blue square symbols), the spin-orbital torque $\langle T_\mathrm{SO}^{S_y} \rangle$ (red circle symbols), and the exchange torque $\langle T_\mathrm{XC}^{S_y} \rangle$ (orange triangle symbols) in each layer as an external electric field is applied to Cr layers (yellow shaded region). 
(b) Each contribution summed over different range of the FM layers.
}
\end{figure}

The intrinsic response of the OAM can interact with the magnetic moment by the SOC, which affects the current-induced torque together with the spin-injection. To analyze the torque, we evaluate the intrinsic responses of the spin flux $\Phi^{Sy}$, \emph{spin-orbital} torque $T_\mathrm{SO}^{S_y}$, and exchange torque $T_\mathrm{XC}^{S_y}$, which appear in the continuity equation for the spin \cite{Haney2010, Go2020b},
$
{\partial \mathbf{S}}/\partial t = 
\Phi^{\mathbf{S}} + T_\mathrm{SO}^\mathbf{S} + T_\mathrm{XC}^\mathbf{S}.
$
Note that $T_\mathrm{XC}^\mathbf{S}$ describes the angular momentum transfer between the spin of the electron and local magnetic moment, which is responsible for the magnetization dynamics. The spin-orbital torque is related to the induced OAM by $T_\mathrm{SO}^{S_y} = \lambda_\mathrm{SO} \mathbf{L} \times \mathbf{S}|_y \sim L_x \hat{\mathbf{x}}\times \hat{\mathbf{z}}$, where $\lambda_\mathrm{SO}$ is the strength of the SOC in CoFe and $\hat{\mathbf{z}}$ is the direction of the magnetization in equilibrium. In the steady state, the dampinglike component of the torque on the magnetization is given by
\begin{eqnarray}
\label{eq:steady_state}
T_\mathrm{DL} 
= - \langle T_\mathrm{XC}^{S_y} \rangle 
\approx 
\langle \Phi^{S_y} \rangle + \langle T_\mathrm{SO}^{S_y} \rangle.
\end{eqnarray}
We denote the first and second terms as the spin and orbital contributions to the torque on the magnetization. 

Figure~\ref{fig:torque_analysis}(a) shows intrinsic response of each term in the spin continuity equation in Cr(20)/CoFe(40). As expected, $\langle \Phi^{S_y} \rangle$ exhibits an oscillatory decaying behavior, where the spin dephasing occurs over $\sim 15$ atomic layers. On the other hand, $\langle T_\mathrm{SO}^{S_y} \rangle$ displays a monotonic decay without any oscillation. Moreover, it persists over longer distance, up to $\sim 30$ atomic layers, with the behavior reminiscent of $\langle L_x \rangle$ in Fig.~\ref{fig:induced_OM}(b). It might appear that $\langle T_\mathrm{XC}^{S_y} \rangle$ is dominated by $\langle \Phi^{S_y} \rangle$ which is an order of magnitude larger than $\langle T_\mathrm{SO}^{S_y} \rangle$. However, a close inspection of the total torque reveals that the orbital contribution may overcome the spin contribution. Figure~\ref{fig:torque_analysis}(b) shows $\langle \Phi^{S_y} \rangle$, $\langle T_\mathrm{SO}^{S_y} \rangle$, and $\langle T_\mathrm{XC}^{S_y} \rangle$ summed over different range of the FM layers from the interface layer. The sum of $\langle \Phi^{S_y} \rangle$ converges to a saturation value in $\sim 15$ atomic layers, which is the spin dephasing length of CoFe. However, the sum of $\langle T_\mathrm{SO}^{S_y} \rangle$ exhibits a slow monotonic increase up to $\sim 30$ atomic layers. As a result, while $\langle T_\mathrm{XC}^{S_y} \rangle$ is dominated by  $\langle \Phi^{S_y} \rangle$ when the summation range is small, $\langle T_\mathrm{SO}^{S_y} \rangle$, i.e. the orbital contribution, becomes more important as the summation range increases. Note that the oscillation of $\langle T_\mathrm{XC}^{S_y} \rangle$ comes from the oscillation of $\langle \Phi^{S_y} \rangle$, but overall negative slope of $\langle T_\mathrm{XC}^{S_y} \rangle$ is due to a positive slope of $\langle T_\mathrm{SO}^{S_y} \rangle$, following Eq.~\eqref{eq:steady_state}. It is remarkable to observe a sign change in $\langle T_\mathrm{XC}^{S_y} \rangle$ as the summation range becomes larger than 10 layers, where the orbital and spin contributions cancel each other.

This result implies that the orbital contribution to the current-induced torque can be dominant over the spin contribution as the thickness of the FM becomes larger than the spin dephasing length. Therefore, measuring the current-induced torque as a function of the FM thickness can be a way to experimentally confirm the OT. Here, a slow saturation behavior of the torque with respect to the FM thickness compared to the spin dephasing length can be a crucial evidence of the effect, as shown by Refs. \cite{Ding2022, Hayashi2022, Liao2022}. Another possible contribution could arise from the FM itself via anomalous spin-orbit torque \cite{Wang2019, Cespedes-Berrocal2021}. However, we find that this contribution exhibits a qualitatively different spatial profile and the saturation length scale is determined by the spin dephasing \cite{Supplementary}.

\begin{figure}[t!]
\includegraphics[angle=0, width=0.45\textwidth]{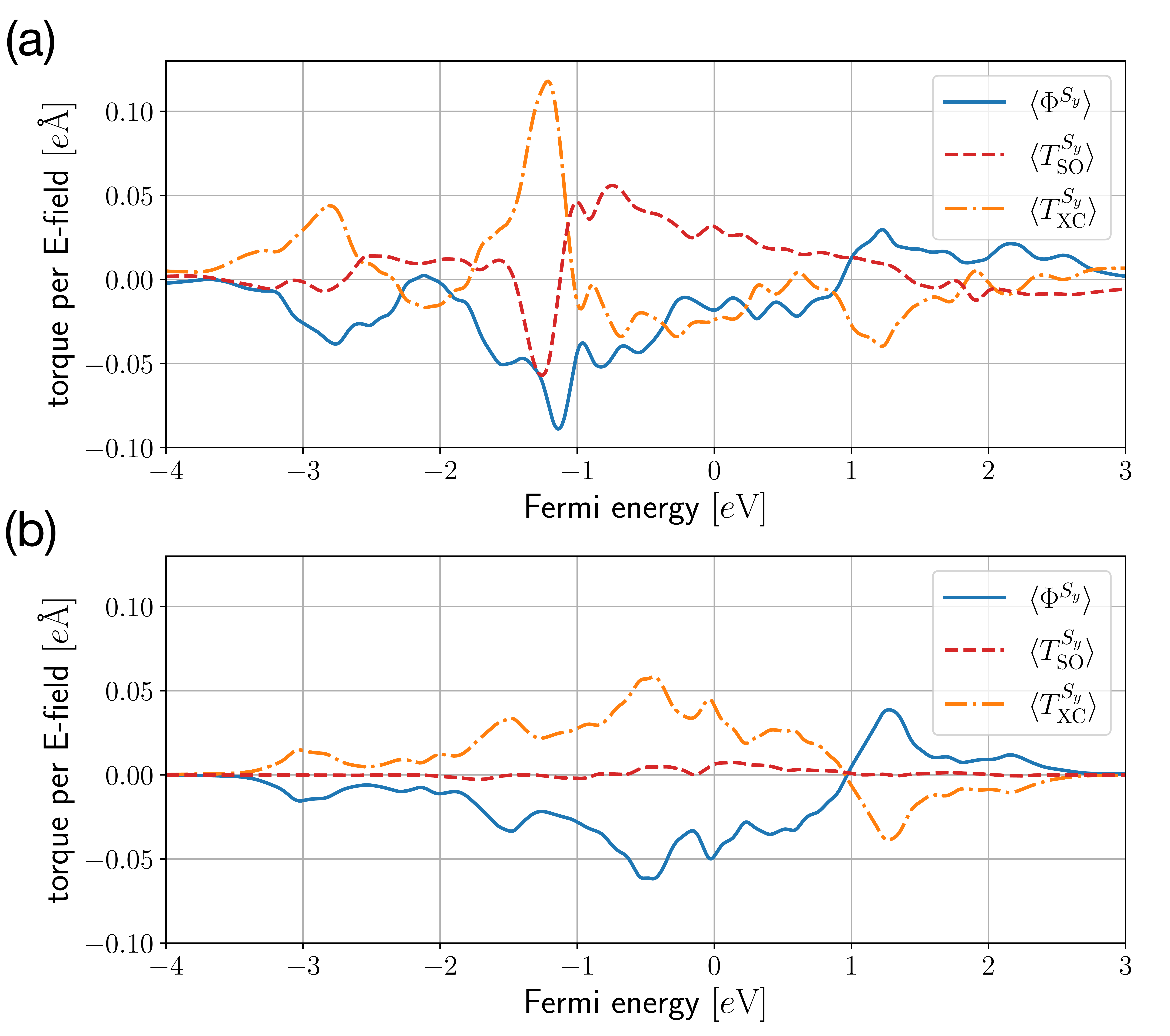}
\caption{
\label{fig:fermi_dep}
(a) Fermi energy dependence of $\langle \Phi^{S_y} \rangle$ (blue solid line), $\langle T_\mathrm{SO}^{S_y} \rangle$ (red dashed line), and $\langle T_\mathrm{XC}^{S_y} \rangle$ (orange dash-dot line), which are summed over all CoFe layers in Cr(20)/CoFe(40). (b) The result for a hypothetical system where the off-diagonal orbital hybridization is neglected in CoFe.
}
\end{figure}


Because the long-range nature of the orbital MEC is due to nearly degenerate orbital characters, removing the orbital hybridization in the FM may suppress $\langle T_\mathrm{SO}^{S_y} \rangle$. In order to demonstrate this point, we setup a hypothetical system where the off-diagonal orbital hybridizations in CoFe are absent, and compare the result with the pristine system. Figure~\ref{fig:fermi_dep}(a) shows Fermi energy dependence of each of the terms in the spin continuity equation for the pristine system, which are summed over all CoFe layers. As shown in Fig.~\ref{fig:torque_analysis}(b), the magnitude of $\langle T_\mathrm{SO}^{S_y} \rangle$ is larger than that of $\langle \Phi^{S_y} \rangle$, and thus $\langle T_\mathrm{XC}^{S_y} \rangle$ is dominated by $\langle T_\mathrm{SO}^{S_y} \rangle$. 
However, in the hypothetical system, $\langle T_\mathrm{SO}^{S_y} \rangle$ is suppressed to nearly zero [Fig.~\ref{fig:fermi_dep}(b)], and $\langle T_\mathrm{XC}^{S_y} \rangle$ has a one-to-one correlation with $\langle \Phi^{S_y} \rangle$, meaning that only the spin injection contributes to the torque on the magnetization. Surprisingly, the overall features of $\langle \Phi^{S_y} \rangle$ are similar in both pristine and hypothetical systems.
This demonstrates that the spin injection is less susceptible to the crystal structure while the orbital injection depends crucially on the crystallinity which determines the orbital hybridization. We note that a recent experiment on AlO$_x$/Cu/FM heterostructures observed a dramatic dependence of the torque efficiency on the interface crystallinity, whose mechanism was attributed to the orbital current \cite{Kim2021}.

The mechanism of the orbital torque has been understood so far as an ``orbital-to-spin'' conversion, but the underlying microscopic mechanism needs a more precise description. In Ref.~\cite{Go2020}, some of us explained that the spin current converted from the orbital current in the FM can exert torque on the magnetization, which appears in the first order in $\lambda_\mathrm{SO}$ of the FM. In fact, this mechanism explains an overall positive slope of $\langle \Phi^{S_y} \rangle$ in Fig.~\ref{fig:torque_analysis}(b), which disappears when the SOC is absent in the FM \cite{Supplementary}. However, the long-range behavior of orbital magnetoelectric torque is additional contribution which has not been noticed so far \cite{Go2020Comment}. Here, $\langle L_x \rangle$ is already first order in $\lambda_\mathrm{SO}$, thus $\langle T_\mathrm{SO}^\mathrm{S_y} \rangle$ appears in the second order in $\lambda_\mathrm{SO}$ \cite{Supplementary}. What is remarkable is that despite small SOC of the FM, $\lambda_\mathrm{SO} \approx 70\ \mathrm{meV}/\hbar^2$ for CoFe, the contribution from $\langle T_\mathrm{SO}^\mathrm{S_y} \rangle$ may overcome that of $\langle \Phi^{S_y}\rangle$ in the current-induced torque. This suggests that  hotspots play an essential role for gigantic response of $\langle L_x \rangle$.

In conclusion, we uncovered nonlocal nature of the orbital magnetoelectric torque in a FM in contact with a NM, which propagates long distance away from the interface. In contrast to the expectation that the orbital quenching would suppress OAM, orbital energy levels imposed by the crystal field are responsible for the hotspot nature of the response, where the dynamics is trapped due to nearly degenerate orbital character. This is a unique feature of the orbital MEC, whose analog does not exist for the spin. We showed that the OAM in the FM monotonically decreases without any oscillation while the spin rapidly oscillates and decays. As a result, the orbital contribution to the torque on the magnetization can be dominant over the spin contribution, especially when the FM is thicker than the spin dephasing length. These findings may prove to be critical for experimental detection of orbital magnetoelectic torque and open a venue toward orbitronic applications.

\begin{acknowledgements}
D.G acknowledges discussion with Tatiana G. Rappoport, Henri Jaffr\`{e}s, Vincent Cros, and Albert Fert. We also thank Hiroki Hayashi and Kazuya Ando for sharing experimental data and providing insightful comments. We gratefully acknowledge the J\"ulich Supercomputing Centre for providing computational resources under project jiff40. This work was funded by the Deutsche Forschungsgemeinschaft (DFG, German Research Foundation) $-$ TRR 173/2 $-$ 268565370 (project A11), TRR 288 $-$ 422213477 (project B06). D.J. was supported by the Global Ph.D. Fellowship Program by National Research Foundation of Korea (Grant No. 2018H1A2A1060270). D.J. and H.-W.L acknowledge the financial support from the Samsung Science and Technology Foundation (Grant No. BA-1501-51). K.-W.K acknowledges the financial support from the National Research Foundation (NRF) of Korea (2020R1C1C1012664) and the KIST Institutional Programs (2E31541,2E31542).
B.-G.P acknowledges financial support from the National Research Foundation of Korea (2020R1A2C2010309).
\end{acknowledgements}

\bibliography{bib_long_range}

\begin{thebibliography}{58}
\bibitem [{Go2()}]{Go2020Comment}%
  \BibitemOpen
  \href@noop {} {}\bibinfo {note} {In the model study of Ref. \cite{Go2020},
  the orbital hybridization in the FM is neglected, which explains why the
  second mechanism presented this Letter is missing.}\BibitemShut {Stop}%
\end{thebibliography}

\end{document}